\documentstyle[twocolumn,prb,aps]{revtex}
\def\beq{\begin{equation}}
\def\eeq{\end{equation}}
\def\beqa{\begin{eqnarray}}
\def\eeqa{\end{eqnarray}}

\begin{document}
\draft
\title{An equation of state {\em \`a 
la} Carnahan-Starling for a five-dimensional fluid of hard hyperspheres}
\author{Andr\'es Santos\cite{email}}
\address{Departamento de F\'{\i}sica,
Universidad de Extremadura,
E-06071 Badajoz, Spain}
\date{\today}
\maketitle
\narrowtext
\begin{abstract}
The equation of state for five-dimensional hard hyperspheres arising as a 
weighted average of the Percus-Yevick compressibility ($\frac{3}{5}$) and
virial ($\frac{2}{5}$) equations of state is considered. This 
Carnahan-Starling-like equation turns out to be extremely accurate, despite 
the fact that both Percus-Yevick equations of state are rather poor.
\end{abstract}

Although not present in nature, fluids of hard hyperspheres in 
high dimensions ($d\geq 4$) have attracted the attention of a number of 
researchers over the last twenty 
years.\cite{FI81,LB82,J82,L84,MT84,BC87,SMS89,ASV89,SM90,LM90,GGS90,MSAV91,GGS91,BMC99}
Among these studies, one of the most important outcomes was the realization 
by Freasier and Isbister\cite{FI81} and, independently, by 
Leutheusser\cite{L84} that the Percus-Yevick (PY) equation\cite{HM86} admits 
an exact solution for a system of hard spheres in $d=\text{odd}$ dimensions.
In the special case of a five-dimensional system ($d=5$), the 
virial series representation of the compressibility factor $Z\equiv p/\rho 
k_B T$ (where $p$ is the pressure, $\rho$ is the number density, $k_B$ is 
the Boltzmann constant, and $T$ is the temperature) is
$Z(\eta)=\sum_{n=0}^\infty b_{n+1} \eta^n$,
where $\eta=(\pi^2/60)\rho\sigma^5$ is the volume fraction ($\sigma$ being 
the diameter of a sphere) and $b_n$ are (reduced) virial 
coefficients. The exact values of the first four  virial 
coefficients are \cite{LB82,LM90} $b_1=1$, $b_2=16$, $b_3=106$, and 
$b_4=311.18341(2)$. The fifth virial coefficient was 
estimated by Monte Carlo integration to be  $b_5\simeq 970$.\cite{FI81}
More recent and accurate Monte Carlo calculations yield 
$b_5=843(4)$ and $b_6=988(28)$.\cite{BMC99}
The exact 
 knowledge of the 
virial coefficients $b_1$--$b_4$ and in some cases of the Monte Carlo 
values for $b_5$ and $b_6$ has been exploited to construct several
approximate equations of state (EOS), several of them being reviewed in 
Ref.\ \onlinecite{BMC99}.

One of the simplest proposals is Song, Mason, and Stratt's (SMS),\cite{SMS89}
who, by viewing the Carnahan-Starling (CS) EOS for $d=3$\cite{CS69} as 
arising from a kind of mean-field theory,  arrived at a generalization for 
$d$ dimensions that makes use of the first three virial coefficients. Baus 
and Colot (BC)\cite{BC87} proposed a {\em rescaled\/} (truncated) virial 
expansion that explicitly accounts for the first four virial coefficients. 
A slightly more sophisticated EOS is the {\em rescaled\/} 
Pad\'e approximant proposed by Maeso {\em et al.} (MSAV),\cite{MSAV91} 
which  reads
\beq
\label{5}
Z_{\text{MSAV}}(\eta)=\frac{1+p_1\eta+p_2\eta^2}{(1-\eta)^5(1+q_1\eta)},
\eeq
where $p_1=(776-b_4)/36$, $p_2=(5476-11b_4)/36$, and $q_1=(380-b_4)/36$.
One of the most accurate proposals to date is the semi-empirical EOS
proposed by Luban and Michels.\cite{LM90} These authors first introduce a 
function $\zeta(\eta)$ defined by \beq
\label{4}
Z_{\text{LM}}(\eta)=1+\frac{b_2\eta\left\{1+\left[b_3/b_2- 
\zeta(\eta)b_4/b_3\right]\eta\right\}}
{1-\zeta(\eta)(b_4/b_3)\eta+\left[\zeta(\eta)-1\right] 
(b_4/b_2)\eta^2}.
\eeq
 Equation (\ref{4}) is consistent with the exact first four virial 
coefficients, regardless of the choice of  $\zeta(\eta)$. The approximation 
$\zeta(\eta)=1$ is equivalent to assuming a  Pad\'e approximant [2,1] for 
$Z(\eta)$. Instead, Luban and Michels observed that the computer simulation 
data of Ref.\ \onlinecite{MT84}, $\{Z_{\text{sim}}(\eta_i), i=1,\ldots,8\} 
$ [cf. Table \ref{table2}], favor a {\em linear\/} approximation for  
$\zeta(\eta)$ and, by a least-square fit, they found
$
\zeta(\eta)=1.074(16)+0.350(96)\eta
$.
Another semi-empirical EOS (not included in Ref.\ 
\onlinecite{BMC99}) was proposed by Amor\'os {\em et al.} (ASV):\cite{ASV89}
\beq
\label{ASV}
Z_{\text{ASV}}(\eta)=\sum_{n=0}^4 
\beta_{n+1}\eta^n+\frac{5\eta_0}{\eta_0-\eta}+C{\eta^4}
\left[\frac{1}{(1-\eta)^4}-1\right].
\eeq
This equation imposes a single pole at the close-packing fraction 
$\eta_0=\sqrt{2}\pi^2/30$. The parameters $\beta_n=b_n-5\eta_0^{-(n-1)}$ 
are fixed so as to 
reproduce the first five virial coefficients,  while $C$ is determined by a 
fit to simulation data.\cite{MT84}
By using the presently known values of 
$b_4$ and $b_5$  and minimizing 
$\sum_{i=1}^8 [1-Z_{\text{ASV}}(\eta_i)/Z_{\text{sim}}(\eta_i)]^2$ one 
finds $C=276.88$.
Finally, Pad\'e approximants [2,3] and [3,2] have also been 
considered.\cite{BMC99}

All the previous EOS rely upon some extra information, 
such as   known virial coefficients and simulation data. On the other 
hand, (approximate) integral equation theories\cite{HM86} provide  the 
correlation functions  describing the structure of the fluid. From these 
functions one can obtain the EOS, that usually adopts a different form 
depending on the route followed. 
As said above, the PY integral equation has an exact solution for 
a system of hard spheres in odd dimensions. In particular, the 
analytical expressions of the EOS obtained from the virial 
route,  $Z_{\text{PY-v}}(\eta)$, and from the 
compressibility route, $Z_{\text{PY-c}}(\eta)$ are 
known for $d=5$.\cite{FI81,L84,GGS90,BMC99}
Nevertheless, these two EOS are highly inconsistent with each 
other.\cite{FI81,BMC99}
This inconsistency problem is also present with lower 
dimensions (except in the one-dimensional case, where the PY theory becomes 
exact), but to a lesser extent.
This led Freasier and Isbister \cite{FI81} to conclude that ``the 
PY approximation for hard cores is an increasingly bad 
approximation as the dimensionality of the system grows larger.''

As is well known, the CS EOS for 
three-dimensional hard spheres\cite{CS69} plays a prominent role in 
liquid state theory.\cite{HM86} 
While originally derived from the observation that the numerical values of 
the known virial coefficients came remarkably close to fitting a simple 
algebraic expression,\cite{CS69} the CS equation is usually viewed as a 
suitable linear combination of the compressibility and virial EOS
resulting from the PY theory,\cite{HM86} namely
\beq
\label{9}
Z_{\text{CS}}(\eta)=\alpha^{(d)} 
Z_{\text{PY-c}}(\eta)+(1-\alpha^{(d)})Z_{\text{PY-v}}(\eta)
\eeq
with $\alpha^{(3)}=\frac{2}{3}$.
Since, as it happened in the case $d=3$, both PY routes keep bracketing  the 
true values in the case $d=5$,\cite{BMC99} it seems 
natural to wonder whether the simple interpolation formula (\ref{9}) 
works in this case as well. 
This question was addressed by Gonz\'alez {\em et al.},\cite{GGS90} who kept 
the value $\alpha^{(5)}=\frac{2}{3}$. 
The main goal of this 
Note is to propose a different choice 
for $\alpha^{(5)}$. 
The virial coefficients corresponding to $Z_{\text{CS}}(\eta)$ are 
$b_n^{\text{CS}}=\alpha^{(d)} b_n^{\text{PY-c}}+(1-\alpha^{(d)})b_n^{\text{PY-v}}$. 
By using the known values of $b_4$--$b_6$ one gets, however, conflictive 
 estimates for the mixing parameter $\alpha^{(5)}$, 
namely  $\alpha^{(5)}\simeq 
(b_4-b_4^{\text{PY-v}})/(b_4^{\text{PY-c}}-b_4^{\text{PY-v}})\simeq 0.68$,
$\alpha^{(5)}\simeq 
(b_5-b_5^{\text{PY-v}})/(b_5^{\text{PY-c}}-b_5^{\text{PY-v}})\simeq
-0.02$, and $\alpha^{(5)}\simeq 
(b_6-b_6^{\text{PY-v}})/(b_6^{\text{PY-c}}-b_6^{\text{PY-v}})\simeq
0.40$.
On the other hand, minimization of $\sum_{i=1}^8 
[1-Z_{\text{CS}}(\eta_i)/Z_{\text{sim}}(\eta_i)]^2$ yields 
$\alpha^{(5)}=0.62$.
By simplicity, here I take
the rational number $\alpha^{(5)}=\frac{3}{5}$ and propose the corresponding 
EOS (\ref{9}).

Table \ref{table2} compares the MSAV, LM, ASV, and CS  EOS with 
available computer simulations.\cite{MT84} This table complements Table II 
of Ref.\ \onlinecite{BMC99}, where $Z_{\text{MSAV}}$, $Z_{\text{ASV}}$, and 
$Z_{\text{CS}}$ (the latter being proposed in this Note) were not included.
In general, the accuracy of the EOS with adjusted virial 
coefficients improves as the degree of complexity increases. More 
specifically, the average relative deviations from the simulation data are, 
from worse to better, as follows: 4.7\% ($Z_{[3,2]}$),
3.7\% ($Z_{\text{SMS}}$), 3.0\% ($Z_{[2,3]}$), 2.3\% ($Z_{\text{BC}}$), 
0.4\% ($Z_{\text{MSAV}}$),  0.17\% ($Z_{\text{LM}}$), and 0.15\% 
($Z_{\text{ASV}}$).
Concerning the two PY EOS, both are quite poor, with average 
relative deviations equal to 7.1\% ($Z_{\text{PY-v}}$) and 4.2\% 
($Z_{\text{PY-c}}$).
The most interesting point, however, is that the CS-like EOS 
(\ref{9}) (with the choice $\alpha^{(5)}=\frac{3}{5}$) presents an excellent 
agreement with the simulation data (the average relative deviation being 
0.3\%), only slightly inferior to that of the semi-empirical EOS 
$Z_{\text{LM}}$ and $Z_{\text{ASV}}$.
This is especially
remarkable if one considers that  only the first three 
virial coefficients of $Z_{\text{CS}}$ are exact, a circumstance also 
occurring in the case of the original CS equation.\cite{CS69}
The choice $\alpha^{(5)}=\frac{2}{3}$,\cite{GGS90} on the other hand, yields 
an average relative deviation of 0.5\%.

Let me conclude with some speculations. It seems interesting to {\em 
conjecture\/} about the existence of possible ``hidden'' regularities  in 
the PY theory for hard hyperspheres explaining the paradox that, although 
the virial and the compressibility EOS strongly deviate from each other, a 
simple linear combination of them {\em might\/} be surprisingly accurate. 
Since the adequate value of the mixing parameter is 
$\alpha^{(3)}=\frac{2}{3}$ for $d=3$ and $\alpha^{(5)}=\frac{3}{5}$ for 
$d=5$, it is then tempting to {\em speculate\/} that its generalization to 
$d$ dimensions might be $\alpha^{(d)}=(d+1)/{2d}$,\cite{note} so that 
$\alpha^{(\infty)}=\frac{1}{2}$ in the limit of high dimensionality, in 
contrast to other proposals.\cite{GGS91} For $d=7$ the above
 implies 
 that, while 
 $Z_{\text{PY-v}}$ and $Z_{\text{PY-c}}$ would dramatically differ, the 
recipe (\ref{9}) with $\alpha^{(7)}=\frac{4}{7}$ {\em could\/} be very close 
to the true EOS. The confirmation or rebuttal of this expectation would 
require the availability of simulation data for $d=7$, which, to the best of 
my knowledge, are absent at present.
 
I would like to thank Drs.\ S. B. Yuste and M. L\'opez de Haro for a 
critical reading of the manuscript and Dr.\ J. R. Solana for providing
some useful references. Partial financial support by
the DGES (Spain) through grant PB97-1501 and by the Junta de 
Extremadura-Fondo Social Europeo through grant IPR99C031 is also gratefully 
acknowledged.
\narrowtext
\begin{table}
\caption{Compressibility factor $Z$ as obtained from  simulation and from 
 Eq.\ (\protect\ref{5}) (MSAV),
 Eq.\ (\protect\ref{4}) (LM), Eq.\ 
 (\protect\ref{ASV}) (ASV),
 and Eq.\ 
(\protect\ref{9}) with $\alpha^{(5)}=\frac{3}{5}$ (CS, this work)}
\label{table2}
\begin{tabular}{cccccc}
$\rho\sigma^5$&Simulation\tablenote{Ref.\ \protect\onlinecite{MT84}}
&
MSAV
&
LM
&
ASV
&
CS 
 \\
\tableline
0.2&
1.653(1)&
1.653&
1.653&
1.653&
1.653 \\
0.4&
2.624(3)&
2.617&
2.618&
2.618&
2.616 \\
0.6&
4.008(6)&
4.003&
4.009&
4.007&
4.000 \\
0.8&
5.997(11)&
5.964&
5.986&
5.979&
5.964 \\
1.0&
8.748(16)&
8.720&
8.770&
8.758&
8.731 \\
1.1&
10.523(20)&
10.488&
10.553&
10.548&
10.510 \\
1.15&
11.589(22)&
11.490&
11.560&
11.561&
11.520  \\
1.18&
12.217(24)&
12.133&
12.204&
12.219&
12.168
\\
\end{tabular}
\end{table}
\narrowtext


\end{document}